\documentclass[12pt]{article}
\usepackage{color}
\usepackage{graphicx}
\def\nabstar#1{\nabla\kern-0.5pt\smash{\raise 4.5pt\hbox{$\ast$}}
               \kern-4.5pt_{#1}}

\def\drvstar#1{\partial\kern-0.5pt\smash{\raise 4.5pt\hbox{$\ast$}}
               \kern-5.0pt_{#1}}

\def\newline{\relax\ifhmode\null\hfil\break\else\nonhmodeerr@\newline\fi}
\def\frac#1#2{{#1\over#2}}
\def\text#1{{\hbox{\rm #1}}}
\def\flushpar{{\par \noindent}}

\newcommand{\beq}{\begin{equation}}
\newcommand{\eeq}{\end{equation}}
\newcommand{\bea}{\begin{eqnarray}}
\newcommand{\eea}{\end{eqnarray}}
\def\Id{ \mbox{1\hspace{-1.2mm}I} }
\def\EQ{\hspace{-2mm} &=& \hspace{-2mm}}

\def\BA{\begin{eqnarray}}
\def\EA{\end{eqnarray}}
\def\BAN{\begin{eqnarray*}}
\def\EAN{\end{eqnarray*}}

\def\nn{\nonumber\\}

\def\tr{\mbox{tr}}

\def\det{\mbox{det}}

\def\g5{\gamma_5}
\def\g4{\gamma_4}
\def\g3{\gamma_3}
\def\g2{\gamma_2}
\def\g1{\gamma_1}

\def\u{{\bf u}}
\def\d{{\bf d}}
\def\s{{\bf s}}
\def\c{{\bf c}}
\def\b{{\bf b}}

\def\cbar{\bar{\bf c}}
\def\bbar{\bar{\bf b}}

\def\Qbar{\bar{\bf Q}}
\input epsf.sty
\newdimen\psfigsize
\def\psfigure#1 #2 #3 #4 #5{
    \begin{figure}[tbh]
      \begin{center}
      \vbox{
        \null\vskip-0.2in\hskip#2
        \epsfxsize=#1
        \epsfbox{#4}
        \vskip -0.3in
        \caption {#5 \label{#3}}
        \vskip 0.0 true in plus 0.3 true in
      }
      \end{center}
   \end{figure}
}
\usepackage{graphicx}
\begin{document}
\thispagestyle{empty}
\begin{flushright}
NTUTH-07-505C \\
May 2007 \\
\end{flushright}
\vskip 2.5truecm
\begin{center}
{\LARGE Beauty mesons in lattice QCD with exact chiral symmetry}
\end{center}
\vskip 1.0truecm
\centerline{{\bf Ting-Wai~Chiu$^{1}$, Tung-Han~Hsieh$^{2}$, 
                 Chao-Hsi~Huang$^{1}$, Kenji~Ogawa$^{1}$}}
\vskip2.0ex
\centerline{$^1\hskip-3pt$ \it
Department of Physics and National Center for Theoretical Sciences} 
\vskip0.5ex
\centerline{\it National Taiwan University, Taipei~10617, Taiwan} 
\vskip2.0ex
\centerline{$^2\hskip-3pt$ \it Research Center for Applied Sciences, 
            Academia Sinica, Taipei~115, Taiwan}
\vskip2.0ex
\centerline{\bf (TWQCD Collaboration)}
\vskip 1cm
\bigskip \nopagebreak \begin{abstract}

\noindent
We present the first study of treating $ \b $, $ \c $, and $ \s $ quarks  
as Dirac fermions in lattice QCD with exact chiral symmetry.
For 100 gauge configurations generated with single-plaquette
action at $ \beta = 7.2 $ on the $ 32^3 \times 60 $ lattice,
we compute point-to-point quark propagators for 33 quark masses
in the range $ 0.01 \le m_q a \le 0.85 $, and measure the
time-correlation function of (pseudo-)scalar, (pseudo-)vector, 
and tensor mesons, for symmetric and asymmetric quark masses respectively.
The lowest-lying mass spectra of mesons
with quark contents $ \b\bbar $, $ \c\bbar $, $\s\bbar $, and $ \c\cbar $  
are determined, together with the pseudoscalar decay constants. 
Our results are sumarized in Tables \ref{tab:cbar_c}-\ref{tab:fP}.
Furthermore, we also determine the $ \b $  and $ \c $ quark masses,     
$ m_{b}^{\overline{\mbox{MS}}}(m_b) = 4.65(5) \mbox{ GeV} $,  
$ m_{c}^{\overline{\mbox{MS}}}(m_c) = 1.16(4) \mbox{ GeV} $.

\vskip 0.8cm
\noindent PACS numbers: 11.15.Ha, 11.30.Rd, 12.38.Gc, 14.40.Lb, 14.40.Nd \\
\noindent Keywords: Lattice QCD, Heavy Quarks, Charmed Mesons, Beauty Mesons

\end{abstract}
\vskip 1.5cm 
\newpage\setcounter{page}1

\section{Introduction}

Currently, one of the most important themes in high energy physics
is to find out whether there is any new physics beyond the Standard Model (SM). 
In order to identify any discrepancies between the high energy experimental 
results and theoretical values derived from the SM, the latter have to be
obtained in a framework which preserves all crucial features of the SM. 
Otherwise, it is difficult to determine whether such a discrepancy 
is due to new physics, or just the approximations (or models) one has used. 

So far, the largest uncertainties in the theoretical predictions
of the SM stem from the sector of the strong interaction, namely, QCD. 
Theoretically, lattice QCD is the most viable framework to tackle QCD 
nonperturbatively from the first principles. However, in practice, 
any lattice QCD calculation suffers from the discretization and finite volume 
errors (which can be systematically improved). 

Since all quarks in QCD are excitations of Dirac fermion fields, 
it is vital to preserve this important feature in any approaches 
to unveil new physics beyond the SM. Theoretically, the most appealing   
lattice fermion scheme is the Domain-Wall/Overlap fermion 
\cite{Kaplan:1992bt,Narayanan:1995gw,Neuberger:1997fp}, which preserves 
the exact chiral symmetry at finite lattice spacing 
\cite{Ginsparg:1981bj}, thus provides a proper formulation of QCD 
on the lattice. However, in practice, it is difficult to accommodate
all quark flavors with presently accessible lattice sizes, since the quark
masses span at least 3 orders of magnitude 
(from $ m_{u} \sim 5 $ MeV to $ m_b \sim 4600 $ MeV), 
even excluding the top quark.  
In fact, one still cannot put the physical $ \u/\d $ quark on a lattice, 
with sufficiently large number of sites in each direction such that 
the discretization and finite volume errors are well under control. 
Thus, one can only perform simulations at unphysically heavy 
$ \u/\d $ quark masses, and then use chiral perturbation theory to 
extrapolate to the physical limit (i.e., $ m_\pi \simeq 140 $ MeV). 
This scenario is not expected to have dramatic changes  
in the next few years.        

Nevertheless, if one only aims at putting $ \b $, $ \c $, and $ \s $ 
quarks on the lattice (with $ m_q a < 1 $, and $ M_{h} L > 4 $), 
then a lattice of size $ \sim 32^3 \times 60 $  
(with inverse lattice spacing $ a^{-1} = 7.68 $ GeV)  
seems to be sufficient for this purpose. 
As we will see below, the meson mass spectra 
turn out in good agreement with 
the experimental values, for quark contents $ \b\bbar $, $ \c\bbar $, 
$ \c\cbar $, and $ \s\bbar $.   

Note that our approach is fundamentally different from other 
lattice QCD calculations using relativistic heavy quark formalism,   
or heavy quark effective theory (HQET), or non-relativistic QCD (NRQCD),  
in which the systematic errors are difficult to control. 
  
In this paper, we will restrict ourselves to mesons with quark
contents $ \b\bbar $, $ \c\bbar $, $ \c\cbar $, and $ \s\bbar $. 
Our results of the masses and decay constants of the pseudoscalar 
mesons $ B_s $ and $ B_c $,
and also the masses of the vector mesons $ B_s^* $ and $ B_c^* $,
have been presented in Ref. \cite{Chiu:2007bc}.
For the mass spectra of spin-1/2 and spin-3/2 baryons 
with quark contents $ \b\b\b $, $ \b\b\c $, $\b\c\c$, 
$ \b\c\s $, $ \b\s\s $, and $ \c\c\c $, our results will be 
presented in a forthcoming paper.

\section{Lattice quarks with exact chiral symmetry}

To implement exact chiral symmetry on the lattice 
\cite{Kaplan:1992bt,Narayanan:1995gw,Neuberger:1997fp,Ginsparg:1981bj}, 
we consider the optimal domain-wall fermion proposed by 
Chiu \cite{Chiu:2002ir,Chiu:2003ir}. 
The action of optimal domain-wall fermion can be written as \cite{Chiu:2003ir}
\BAN
\label{eq:ODWF}
{\cal A}_F \EQ 
\sum_{s,s'=0}^{N_s+1} \sum_{x,x'}
\bar\psi(x,s)
\{ (\omega_s D_w(x,x') + \delta_{x,x'}) \delta_{ss'} \nn
&& \hspace{12mm} + (\omega_s D_w(x,x') - \delta_{x,x'})
    (P_{+} \delta_{s',s-1} + P_{-} \delta_{s',s+1} ) \} \psi_(x',s')
\EAN
with boundary conditions
\BAN
\label{eq:bc1a}
P_{+} \psi(x,-1) \EQ -r \ m_q \ P_{+} \psi(x,N_s+1), \\
\label{eq:bc2a}
P_{-} \psi(x,N_s+2) \EQ -r \ m_q \ P_{-} \psi(x,0), 
\hspace{4mm} r=\frac{1}{2m_0}, 
\EAN
where $ m_q $ is the bare quark mass, and the weights 
$ \{ \omega_s, s = 1, \cdots, N_s \} $  
are specified by the exact formula derived in Ref. \cite{Chiu:2002ir} 
such that the system possesses the maximal chiral symmetry for any fixed 
$ N_s $. Here $ H_w = \gamma_5 D_w $, and $ D_w $ is the standard Wilson 
Dirac operator plus a negative parameter $ -m_0 $ ($ 0 < m_0 < 2 $). 
The quark fields are
constructed from the boundary modes at $ s=0 $ and $ s=N_s + 1 $
with $ \omega_0 = \omega_{N_s + 1} = 0 $ \cite{Chiu:2003ir}:
\BAN
\label{eq:q}
q(x) &=& \sqrt{r} \left[ P_{-} \psi(x,0) + P_{+} \psi(x,N_s+1) \right], \\ 
\label{eq:qbar}
\bar q(x) &=& \sqrt{r}
\left[ \bar\psi(x,0) P_{+} + \bar\psi(x,N_s+1) P_{-} \right].
\EAN
After introducing pseudofermions with $ m_q = 2 m_0 $, 
the generating functional for $n$-point Green's function
of the quark fields can be derived as \cite{Chiu:2003ir}, 
\bea
\label{eq:ZW_odwf}
Z[J,\bar J] =
\frac{\int [dU] e^{-{\cal A}_g} \det [(D_c+m_q)(1+rD_c)^{-1}]
           \exp \left\{ \bar J (D_c+ m_q)^{-1} J \right\}  }
     {\int [dU] e^{-{\cal A}_g} \det [(D_c + m_q)(1+r D_c)^{-1}] }
\eea
where $ {\cal A}_g $ is the action of the gauge fields,
$ \bar J $ and $ J $
are the Grassman sources of $ q $ and $ \bar q $ respectively, and
\BAN
\label{eq:Dc}
D_c \EQ 2 m_0 \frac{ 1 + \gamma_5 S_{opt} }{ 1 - \gamma_5 S_{opt} } \ , \\
S_{opt} \EQ \frac{1 - \prod_{s=1}^{N_s} T_s}
                {1 + \prod_{s=1}^{N_s} T_s}, \\
 T_s \EQ \frac{1 - \omega_s H_w }{1 + \omega_s H_w} . 
\EAN

From (\ref{eq:ZW_odwf}), the valence quark propagator 
in background gauge field is
\BAN
\label{eq:quark_prop}
\langle q(x) \bar q(y) \rangle \EQ
- \left. \frac{\delta^2 Z[J,\bar J]}{\delta \bar J(x) \delta J(y)}
\right|_{J=\bar J=0}                                             
= (D_c + m_q)^{-1}_{x,y}
\EAN
where $ D_c $ is exactly chirally symmetric 
($ D_c \gamma_5 + \gamma_5 D_c = 0 $) 
in the limit $ N_s \to \infty $, and its deviation from exact chiral 
symmetry due to finite $ N_s $ is the {\it minimal} provided that  
the weights $ \{ \omega_s \} $ are fixed according to the formula 
derived in Ref. \cite{Chiu:2002ir}. Note that in this framework,  
the bare mass $ m_q $ (no matter heavy or light) 
in the valence quark propagator $ (D_c + m_q)^{-1} $ 
is well-defined for any gauge configurations.

We generate 100 gauge configurations with single plaquette gauge action
at $ \beta = 7.2 $ on the $ 32^3 \times 60 $ lattice.
For $ m_0 = 1.3 $ and $ N_s = 128 $, we fix 
the weights $ \{ \omega_s \} $ with $ \lambda_{min} = 0.1 $ 
and $ \lambda_{max} = 6.4 $, 
where $ \lambda_{min} \le \lambda(|H_w|) \le \lambda_{max} $
for all gauge configurations.    
For each configuration, point to point valence quark propagators are 
computed for 33 bare quark masses in the range $ 0.01 \le m_q a \le 0.85 $, 
with stopping criteria $ 10^{-11} $ for the conjugate gradient.
Then the norm of the residual vector of each column of the quark propagator 
is less than $ 2 \times 10^{-11} $  
\BAN
|| (D_c + m_q ) Y - \Id || < 2 \times 10^{-11},  
\EAN
and the chiral symmetry breaking due to finite $ N_s $ is 
less than $ 10^{-14} $,
\BAN
\sigma = \left| \frac{Y^{\dagger} S_{opt}^2 Y}{Y^{\dagger} Y} - 1 \right|
< 10^{-14},
\EAN

In this paper, we measure the time-correlation function 
\bea
\label{eq:C}
C_{\Gamma} (t) \EQ
\left< 
\sum_{\vec{x}}
\tr\{ \Gamma (D_c + m_Q)^{-1}_{x,0} \Gamma (D_c + m_q)^{-1}_{0,x} \} 
\right>  
\eea
for scalar ($S$), pseudoscalar ($P$), vector ($V$), axial-vector ($A$), 
and tensor ($T$) mesons, with Dirac matrix
$\Gamma=\{\Id,\gamma_5,\gamma_i,\gamma_5\gamma_i,\gamma_5\gamma_4\gamma_i \} $
respectively. For vector mesons, we average over $i=1,2,3$ components, 
namely,   
\BAN
\label{eq:CV}
C_V (t) = \left< 
\frac{1}{3} \sum_{i=1}^3 \sum_{\vec{x}}
\tr\{ \gamma_i (D_c + m_Q)^{-1}_{x,0} 
      \gamma_i (D_c + m_q)^{-1}_{0,x} \} \right>
\EAN
Similarly, we perform the same averaging for axial-vector and tensor mesons. 

The time-correlation function $ C_{\Gamma}(t) $ is measured 
for the following three categories: 
\begin{itemize}
\item Symmetric masses with $ m_Q = m_q $ for 33 quark masses. 
\item Asymmetric masses with fixed $ m_Q = m_b = 0.68 a^{-1} $,
      and $ m_q $ running over all different quark masses.      
\item Asymmetric masses with fixed $ m_Q = m_c = 0.16 a^{-1} $,
      and $ m_q $ running over all different quark masses.      
\end{itemize}

\section{Determination of $ a^{-1} $, $ m_b $, $ m_c $, and $ m_s $}

In Ref. \cite{Chiu:2005ue}, we determine the inverse lattice 
spacing from the pion decay constant, with experimental input 
$ f_\pi = 131 $ MeV. However, in this paper, we do not use
the same method since the smallest quark mass   
turns out to be rather heavy ($ \simeq m_s/2 $), thus chiral 
extrapolation to $ m_q \simeq 0 $ does not seem to be feasible. 
Nevertheless, we can use the mass and decay constant of the 
pseudoscalar meson $ \eta_c(2980) $ to determine $ m_c $ and 
$ a^{-1} $ simultaneously. This can be seen as follows.   

For symmetric masses $ m_Q = m_q $, the pseudoscalar
time-correlation function $ C_{P} (t) $ ($ \Gamma = \gamma_5 $) 
is measured, and fitted to the usual formula
\BAN
\label{eq:Gt_fit}
\frac{z^2}{2 m_{P} a } [ e^{-m_{P} a t} + e^{-m_{P} a (T-t)} ]
\EAN
to extract the mass $ m_{P} a $ and the decay constant
\BAN
\label{eq:fpi}
f_{P} a = 2 m_q a \frac{z}{m_{P}^2 a^2 } \ .
\EAN
Then the ratio $ m_{P}/f_{P} $ can be obtained for each $ m_q $.

Since $ f_{\eta_c} $ has not been measured in HEP experiments, we 
do not have a physical value for the ratio $ m_{\eta_c}/f_{\eta_c} $.  
Nevertheless, we can obtain the ratio $ m_{\eta_c}/f_{\eta_c} = 6.8(2) $
from our previous study of pseudoscalar mesons on the 
$ 20^3 \times 40 $ lattice at $ \beta = 6.1 $ \cite{Chiu:2005ue}. 
Then we can use this theoretical value 6.8 to discriminate
which $ m_q $ can give the ratio $ m_{P}/f_{P} $ closest to this value. 
We find that at $ m_q a = 0.16 $ the ratio $ m_{P}/f_{P} = 6.8(1) $, 
which is the closest to 6.8. Thus we fix $ m_c a = 0.16 $.
Then we use the experimental mass of $ \eta_c(2980) $ to determine
$ a^{-1} $ through the relation  
$$ 
m_{P} a|_{m_c} = (2980 \mbox{ MeV}) \times a = 0.388(3) 
$$
and obtain $ a^{-1} = 7680(59) $ MeV. 
To check the goodness of the values of $ m_c $ and $ a^{-1} $, 
we compute the time-correlation function of $ \cbar \gamma_i \c $, 
and extract the mass of the vector meson equal to $ 3091(11) $ MeV, 
in good agreement with $ J/\Psi(3097) $. 

The bare mass of strange quark is determined by extracting the
mass of vector meson from the time-correlation function
$ C_V(t) $.  At $ m_q a = 0.02 $, $ m_V a = 0.1337(5) $,
which gives $ m_V = 1027(38) $ MeV, in good agreement with
the mass of $ \phi(1020) $. Thus we take the strange quark
bare mass to be $ m_s a = 0.02 $.
Similarly, at $ m_q a = 0.68 $, $ m_V a = 1.2308(4) $,
which gives $ m_V = 9453(3) $ MeV, in
good agreement with the mass of $ \Upsilon(9460) $.
Thus, we fix the bottom quark bare mass to be $ m_b a = 0.68 $.

Note that the spatial size of our lattice ($ L \simeq 0.8 $ fm) 
seems to be small at first glance, however, even for the smallest 
quark mass $ m_q a = 0.01 $, its pseudoscalar mass satisfies 
$ m_{P} L > 4 $, thus the finite size effects should be well 
under control.

\section{Charmonium $ \c\cbar $ and Bottomonium $ \b\bbar $}

First of all, we check to what extent we can reproduce the 
charmonium mass spectra which have been measured precisely
by high energy experiments. 

Our results of the mass spectra of the lowest-lying states of 
charmonium are summarized in Table \ref{tab:cbar_c}.  
The first column is the Dirac matrix used for  
computing the time-correlation function (\ref{eq:C}). 
The second column is $ J^{PC} $ of the state. 
The third column is the conventional spectroscopic notation. 
The fourth column is the $ [t_{min}, t_{max}] $ used for fitting  
the data of $ C_\Gamma(t) $ to the usual formula   
$$
\frac{z^2}{2 M a} [ e^{-M a t} + e^{- M a(T-t)} ]  
$$
to extract the meson mass $ M $ and the decay amplitude $ z $.
The fifth column is the mass $ M $ of the state, where the first 
error is statistical, and the second is our estimate of 
systematic error based on all fittings satisfying 
$ \chi^2/\mbox{dof} < 1.3 $ and $ |t_{max} - t_{min}| \ge 6 $ with
$ t_{min} \ge 10 $ and $ t_{max} \le 50 $. 
The last column is the corresponding state in high energy experiments,  
with the PDG mass value \cite{Yao:2006px}.
Evidently, our mass spectra of charmonium are in good agreement 
with the PDG values. 
Note that our result of the hyperfine splitting ($ 1^3 S_1 - 1^1 S_0 $) is   
$ 111(14)(18) $ MeV, comparing with the PDG value $ 118 $ MeV.

For the pseudoscalar $ \eta_c $, we also obtain its decay constant
$ f_{\eta_c} $ together with its mass, through the decay amplitude $ z $ 
in the equation 
\BAN
f_{\eta_c} a = 2 m_c a \frac{z}{m_{\eta_c}^2 a^2 }
\EAN
Our result is 
\bea
\label{eq:f_etac}
f_{\eta_c} = 438 \pm 5 \pm 6 \mbox{ MeV}
\eea
where the first error is statistical, and the second is our estimate of 
systematic error based on all fittings satisfying 
$ \chi^2/\mbox{dof} < 1.3 $ and $ |t_{max} - t_{min}| \ge 6 $ with 
$ t_{min} \ge 10 $ and $ t_{max} \le 50 $.
So far, $ f_{\eta_c} $ has not been determined in high energy experiments.

\begin{table}
\begin{center}
\caption{The mass spectra of lowest-lying ($n=1$) charmonium $\cbar\Gamma\c$ 
         states obtained in this work, in comparison with the PDG values 
         in the last column.}
\vspace{0.5cm}
\begin{tabular}{c|c|c|c|c|c||c}
$ \Gamma $ & $ J^{PC} $ & n$^{2S+1} L_J $ 
           & $ [t_{min},t_{max}] $ & $\chi^2$/dof & Mass(MeV) & PDG \\
\hline
\hline
$ \Id $ & $ 0^{++} $ & $ 1^3 P_0 $ 
                     & [19,42] & 0.32 
                     & 3413(14)(9) 
                     & $ \chi_{c0}(3415) $ \\
$ \gamma_5 $ & $ 0^{-+} $ 
                          & $ 1^1 S_0 $ 
                          & [22,38] & 1.02 
                          & 2980(10)(12) 
                          & $ \eta_c(2980) $ \\ 
$ \gamma_i $ & $ 1^{--} $ & $ 1^3 S_1 $ 
                          & [19,38] & 1.18 
                          & 3091(11)(14) 
                          & $ J/\psi(3097) $ \\
$ \gamma_5\gamma_i $ & $ 1^{++} $ & $ 1^3 P_1 $ 
                                  & [18,42] & 0.51 
                                  & 3516(13)(8) 
                                  & $ \chi_{c1}(3510) $ \\
$ \gamma_5\gamma_4\gamma_i $ & $ 1^{+-} $ & $ 1^1 P_1 $ 
                             & [19,43] & 0.41 
                             & 3526(13)(9) 
                             & $h_c(3524)$ \\
\hline
\end{tabular}
\label{tab:cbar_c}
\end{center}
\end{table}

Next, we turn to the bottomonium ($\b\bbar$) states. 
Our results of the mass spectra of the lowest-lying states of 
bottomonium are summarized in Table \ref{tab:bbar_b}.  

First, we look at the pseudoscalar $ \eta_b $. It was first 
reported by ALEPH Collaboration \cite{Heister:2002if}.
However, it has not been confirmed by other HEP experimental groups.
Thus we suspect that its mass $ 9300(20)(20) $ MeV might not have been 
determined accurately. It is interesting to see whether   
the mass of $ \eta_b $ will turn out to agree with our 
theoretical value $ 9380 \pm 10 $ MeV. 
Besides the mass of $ \eta_b $, we also determined its decay constant 
\bea
\label{eq:f_etab}
f_{\eta_b} = 801 \pm 7 \pm 5 \mbox{ MeV}
\eea
where the first error is statistical, and the second is our estimate of 
systematic error based on all fittings satisfying 
$ \chi^2/\mbox{dof} < 1.3 $ and $ |t_{max} - t_{min}| \ge 6 $ with 
$ t_{min} \ge 10 $ and $ t_{max} \le 50 $.

Finally, we note that the tensor meson $ h_b $ has not been observed 
in high energy experiments, thus our result of its mass $ 9916 \pm 30 $ MeV
serves as the first prediction from lattice QCD with exact chiral symmetry. 
Even though it is obtained in the quenched approximation, 
we suspect that it might provide a reliable prediction of 
hadron mass spectra, especially for mesons with heavy $ \b $ quark.

\begin{table}
\begin{center}
\caption{The mass spectra of lowest-lying ($n=1$) bottomonium $\bbar\Gamma\b$ 
         states obtained in this work. The last column is the   
         experimental state we have identified, and its PDG mass value.}
\vspace{0.5cm}
\begin{tabular}{c|c|c|c|c|c||c}
$ \Gamma $ & $ J^{PC} $ & n$^{2S+1} L_J $ 
           & $ [t_{min},t_{max}] $ & $\chi^2$/dof & Mass(MeV) & PDG \\
\hline
\hline
%
%
$ \Id $ & $ 0^{++} $ & $ 1^3 P_0 $ 
                     & [21,39] & 0.21 
                     & 9863(15)(8) 
                     & $ \chi_{b0}(9859) $ \\
%
$\gamma_5$ & $ 0^{-+} $ & $ 1^1 S_0 $ 
                        & [27,35] & 0.72 
                        & 9383(4)(2) 
                        & $ \eta_b(9300) $ ? \\
%
%
%
%
%
$\gamma_i$ & $ 1^{--} $ & $ 1^3 S_1 $ 
                        & [20,39] & 1.19 
                        & 9453(3)(2) 
                        & $ \Upsilon(9460) $ \\
%
%
%
%
$\gamma_5\gamma_i$ & $ 1^{++}$ & $1^3 P_1$ 
                               & [22,38] & 0.13 
                               & 9896(20)(8) 
                               & $ \chi_{b1}(9893) $ \\
%
%
$\gamma_5\gamma_4\gamma_i$ & $ 1^{+-} $ & $ 1^1 P_1 $ 
                           & [22,38] & 0.10 
                           & 9916(22)(8) 
                           &  \\
\hline 
\end{tabular}
\label{tab:bbar_b}
\end{center}
\end{table}

\section{Mesons with quark contents $ \s\bbar $ and $ \c\bbar $}

The decay constants of heavy-light pseudoscalar mesons (e.g., 
$ f_B $, $ f_{B_s} $, $ f_D $, and $ f_{D_s} $) play an important role 
in extracting the CKM matrix elements which are crucial for 
testing the Standard Model via the unitarity of CKM matrix. 
Theoretically, lattice QCD with exact chiral symmetry provides a
reliable framework to compute the masses and decay constants of
heavy-light pseudoscalar mesons nonperturbatively
from the first principles of QCD.
Our theoretical predictions of $ f_D $ and $ f_{D_s} $ have been 
presented in Ref. \cite{Chiu:2005ue}, 
which turn out in good agreement with the recent  
experimental results \cite{Artuso:2005ym,Artuso:2007zg}
from CLEO Collaboration. For a recent review of heavy-light 
pseudoscalar decay constants from lattice QCD,  
see Ref. \cite{Onogi:2006km} and references therein. 

Before we present our results of $ f_{B_s} $ and $ m_{B_s} $, we
recall the basic formulas as follows.  
In general, the decay constant $ f_P $ of a heavy-light 
pseudoscalar meson $ P $ is defined as  
\BAN
\left<0| A_\mu(0) | P(\vec{p}) \right>  = i p_\mu f_P
\EAN
where $ A_\mu(x) = \bar Q(x) \gamma_\mu \gamma_5 q(x) $ 
is the axial-vector current. Using the formula
$$ \partial_\mu A_\mu = (m_q + m_Q) \bar Q \gamma_5 q $$
one obtains
\BAN
\label{eq:fP}
f_P = (m_q + m_Q )
\frac{| \langle 0| \bar Q \gamma_5 q | P(\vec{0}) \rangle |}{m_P^2}
\EAN
where the pseudoscalar mass $ m_P a $ and the decay amplitude
$ z \equiv | \langle 0| \bar Q \gamma_5 q | P(\vec{0}) \rangle | $
can be obtained by fitting the pseudoscalar time-correlation function
$ C_P(t) $ to the usual formula
$$
\frac{z^2}{2 m_P a} [ e^{-m_P a t} + e^{- m_P a(T-t)} ]
$$

Our results for $ B_s $ are: 
\bea
\label{eq:mf_Bs}
m_{B_s} = 5385 \pm 27 \pm 17 \mbox{ MeV}, \hspace{6mm} 
f_{B_s} = 253 \pm 8 \pm 7 \mbox{ MeV}
\eea
where the first error is statistical, and the second is our estimate of 
systematic error based on all fittings satisfying 
$ \chi^2/\mbox{dof} < 1.3 $ and $ |t_{max} - t_{min}| \ge 6 $ with  
$ t_{min} \ge 10 $ and $ t_{max} \le 50 $.
Our result of $ m_{B_s} $ is in good agreement 
with the experimental value 5368 MeV compiled by PDG. 
Since $ f_{B_s} $ has not been measured in high energy
experiments, our result serves as the first prediction from 
lattice QCD with exact chiral symmetry.

Next, we present our results of the mass spectra of the 
lowest-lying states of beauty mesons with quark content $ \s\bbar $,   
which are summarized in Table \ref{tab:bbar_s}. 
Here we have identified the scalar $ \bbar\s $ meson  
with the state $ B_{sJ}^*(5850) $ observed in HEP experiments, 
due to the proximity of their masses.  
Theoretically, this implies that $ B_{sJ}^*(5850) $ possesses
$ J^P = 0^+ $, which can be verified by HEP experiments 
in the future.  
Moreover, we have obtained the masses of the axial-vector and 
tensor mesons which have not been observed in HEP experiments, 
as a prediction from lattice QCD.

\begin{table}
\begin{center}
\caption{The mass spectra of lowest-lying $ \bbar\Gamma\s$ meson
         states obtained in this work. The last column is the   
         experimental state we have identified, and its PDG mass value.}
\vspace{0.5cm}
\begin{tabular}{c|c|c|c|c|c||c}
$ \Gamma $ & $ J^{P} $ & n$^{2S+1} L_J $ 
           & $ [t_{min},t_{max}] $ & $\chi^2$/dof & Mass(MeV) & PDG \\
\hline
\hline
%
%
$ \Id $ & $ 0^{+} $ & $ 1^3 P_0 $ 
                    & [20,40]  & 0.38 
                    & 5852(15)(12) 
                    & $ B_{sJ}^* (5850) $  \\
%
$\gamma_5$ & $0^{-}$ & $ 1^1 S_0 $  
           & [26,32] & 0.56 
           & 5385(27)(17) 
           & $ B_s(5368) $ \\ 
%
%
%
%
%
$\gamma_i$ & $ 1^{-} $ & $ 1^3 S_1 $ 
           & [25,33] & 0.50 
           & 5424(28)(19) 
           & $ B_s^*(5412)$ \\ 
%
%
%
%
$\gamma_5\gamma_i$ & $ 1^{+} $ & $1^3 P_1$ 
                               & [18,42] & 0.68 
                               & 5884(16)(13) 
                               &  \\
%
%
$\gamma_5\gamma_4\gamma_i$ & $ 1^{+} $ & $ 1^1 P_1 $ 
                           & [18,38] & 0.62 
                           & 5897(16)(12) 
                           &    \\
\hline 
\end{tabular}
\label{tab:bbar_s}
\end{center}
\end{table}

Next, we turn to the heavy mesons with beauty and charm. 
For the pseudoscalar $ B_c $, we obtain  
\bea
\label{eq:mf_Bc}
m_{B_c} = 6278 \pm 6 \pm 4 \mbox{ MeV}, \hspace{6mm} 
f_{B_c} = 489 \pm 4 \pm 3 \mbox{ MeV}
\eea
where the first error is statistical, and the second is our estimate of 
systematic error based on all fittings satisfying 
$ \chi^2/\mbox{dof} < 1.3 $ and $ |t_{max} - t_{min}| \ge 6 $ with 
$ t_{min} \ge 10 $ and $ t_{max} \le 50 $.
Our result of $ m_{B_c} $ is in good agreement with the experimental 
value 6286(5) MeV measured by CDF Collaboration \cite{Abulencia:2005us}, 
while $ f_{B_c} $ has not been measured by HEP experiements. 
In principle, $ f_{B_c} $ can be measured from 
the leptonic decay $ B_c^+ \to l^+ \nu_l $,
since its decay width is proportional to $ f_{B_c}^2 |V_{cb}|^2 $. 

In Table \ref{tab:bbar_c},  
we summarize our results of the mass spectra of the 
lowest-lying states of mesons with beauty and charm. 
Except for the pseudoscalar $ B_c $, other states have not been
observed in experiments. It will be interesting to see to what
extent the experimental results would agree with our theoretical 
values.

\begin{table}
\begin{center}
\caption{The mass spectra of lowest-lying $ \bbar\Gamma\c$ meson
         states obtained in this work. The last column is the   
         experimental state we have identified, and its PDG mass value.}
\vspace{0.5cm}
\begin{tabular}{c|c|c|c|c|c||c}
$ \Gamma $ & $ J^{P} $ & n$^{2S+1} L_J $ 
           & $ [t_{min},t_{max}] $ & $\chi^2$/dof & Mass(MeV) & PDG \\
\hline
\hline
%
%
$ \Id $ & $ 0^{+} $ & $ 1^3 P_0 $ 
                    & [19,41] & 0.26 
                    & 6732(13)(9) 
                    &  \\
%
$\gamma_5$ & $ 0^{-} $ & $ 1^1 S_0 $ 
                       & [19,38] & 1.14 
                       & 6278(6)(4) 
                       & $ B_c(6286) $ \\
%
%
%
%
%
$\gamma_i$ & $ 1^{-} $ & $ 1^3 S_1 $ 
                       & [19,38] & 1.31 
                       & 6315(6)(5) 
                       & \\ 
%
%
%
%
$\gamma_5\gamma_i$ &$ 1^{+}$ & $ 1^3 P_1$ 
                             & [19,41] & 0.27  
                             & 6778(12)(7) 
                             &   \\
%
%
$\gamma_5\gamma_4\gamma_i$ & $ 1^{+} $ & $ 1^1 P_1 $ 
                           & [18,42] & 0.45 
                           & 6796(10)(7) 
                           &    \\
\hline 
\end{tabular}
\label{tab:bbar_c}
\end{center}
\end{table}


\begin{table}
\begin{center}
\caption{The decay constants of pseudoscalar mesons obtained in this work, 
together with their masses. They are identified with the 
corresponding PDG mesons listed in the last column,  
however, the decay constants have not been measured in HEP experiments.}
\vspace{0.5cm}
\begin{tabular}{c|c|c|c|c||c}
$\Qbar \Gamma q $ & $ [t_{min},t_{max}] $ & $\chi^2$/dof 
                  & Mass(MeV) & $f_P$(MeV) & PDG \\
\hline
\hline
$ \bbar\gamma_5\b $ & [27,35] & 0.72 & 9383(4)(2)   & 801(7)(5)  & $ \eta_b(9300) $ \\
$ \bbar\gamma_5\c $ & [19,38] & 1.14 & 6278(6)(4)   & 489(4)(3)  & $ B_c(6287) $ \\
$ \bbar\gamma_5\s $ & [26,32] & 0.56 & 5385(27)(17) & 253(8)(7)  & $ B_s(5368) $ \\ 
$ \cbar\gamma_5\c $ & [22,38] & 1.02 & 2980(10)(12) & 438(5)(6)  & $ \eta_c(2980) $ \\ 
\hline
\end{tabular}
\label{tab:fP}
\end{center}
\end{table}

\section{Masses of $ \b $ and $ \c $ quarks}

In the Standard Model, the quark masses are fundamental parameters which
have to be determined from high energy experiments. However, they cannot
be measured directly since quarks are confined inside hardrons, unlike
an isolated electron whose mass and charge both can be measured directly
from its responses in electric and magnetic fields.
Therefore, the quark masses can only be determined by comparing
a theoretical calculation of physical observables (e.g., hadron masses)
with the experimental values. Evidently, for any field theoretic
calculation, the quark masses depend on the regularization,
as well as the renormalization scheme and scale.
One of the objectives of lattice QCD is to compute the hadron masses
nonperturbatively from the first principles, and from which
the quark masses are determined.

We have used the mass of the vector meson $ \Upsilon(9460) $
to fix the bare mass of $ \b $ quark equal to 
$ m_b = 0.68 a^{-1} = 5.22(4) $ GeV\footnote{Note that we have not
tuned the value of $ m_b $, thus the mass of $ \Upsilon $ 
is 9453(3)(2) MeV (see Table \ref{tab:bbar_b}) rather than 9460 MeV.}   
To transcribe the bare mass 
to the corresponding value in the usual renormalization scheme
$ \overline{\mbox{MS}} $ in high energy
phenomenology, one needs to compute the lattice renormalization
constant $ Z_m = Z_s^{-1} $, where $ Z_s $ is the renormalization
constant for $ \bar\psi \psi $. In general, $ Z_m $ should be
determined nonperturbatively. However, in this case, 
the lattice spacing is rather small ($ a \simeq 0.026 $ fm), 
thus we suspect that the one-loop perturbation 
formula \cite{Alexandrou:2000kj}
\bea
\label{eq:Zs}
Z_s(\mu) = 1 + \frac{ g^2 }{ 4 \pi^2 }
\left[ \mbox{ln} ( a^2 \mu^2 ) + 0.17154 \right], \hspace{8mm}
(m_0=1.30)
\eea
already provides a very good approximation for $ Z_s $.
At $ \beta = 7.2 $, $ a^{-1} = 7.680(59) $ GeV, 
and $ \mu = 2 $ GeV, (\ref{eq:Zs})
gives $ Z_s = 1.086(2) $, which transcribes the bare mass 
$ m_b = 5.22(4) $ GeV to
\BAN
m_{b}^{\overline{\mbox{MS}}}(2 \mbox{ GeV})
= 4.81 \pm 0.04 \mbox{ GeV}  
\EAN
where the error is due to the uncertainty in the lattice spacing. 
Now if we also include the uncertainty in the determination of the 
$ \b $ quark bare mass, which is estimated to be $ \delta(m_b a) = 0.002 $, 
then we obtain 
\bea
\label{eq:mb_MS_2}
m_{b}^{\overline{\mbox{MS}}}(2 \mbox{ GeV})
= 4.81 \pm 0.05 \mbox{ GeV}  
\eea

In order to compare our result with the PDG average,  
we have to obtain $ m_{b}^{\overline{\mbox{MS}}} $ 
at the scale $ \mu = m_b $. This can be obtained by 
solving the equation $ \bar m_b = m_b Z_m(\mu=\bar m_b) $.
Our result is  
\bea
\label{eq:mb_MS_mb}
m_{b}^{\overline{\mbox{MS}}}(m_b) = 4.65 \pm 0.05 \mbox{ GeV}  
\eea
which seems to be higher than the PDG average $ 4.20 \pm 0.07 $
\cite{Yao:2006px}.
   
Now we turn to the $ \c $ quark mass.
Using (\ref{eq:Zs}), the $ \c $ quark bare mass  
$ m_c = 0.160(5) a^{-1} $ is transcribed to 
\bea
\label{eq:mc_MS_2}
m_{c}^{\overline{\mbox{MS}}}(2 \mbox{ GeV}) = 1.13 \pm 0.04 \mbox{ GeV}  
\eea
where the error incorporates the uncertainties in the lattice spacing and 
the $ \c $ quark bare mass.
Again, to compare our result with the PDG average, we obtain 
$ m_{c}^{\overline{\mbox{MS}}} $ at the scale $ \mu = m_c $,  
\bea
\label{eq:mc_MS_mc}
m_{c}^{\overline{\mbox{MS}}}(m_c) = 1.16 \pm 0.04 \mbox{ GeV}  
\eea
which is in good agreement with the PDG average $ 1.25 \pm 0.09 $ 
\cite{Yao:2006px}.

\section{Concluding remark}

We have performed the first study of treating 
$ \b $, $ \c $, and $ \s $ quarks  
as Dirac fermions in lattice QCD with exact chiral symmetry.
The lowest-lying mass spectra of mesons
with quark contents $ \b\bbar $, $ \c\bbar $, $\s\bbar $, and $ \c\cbar $  
are determined, together with the pseudoscalar decay constants. 
Our results of the meson mass spectra 
in Tables \ref{tab:cbar_c}-\ref{tab:bbar_c}, pseudoscalar  
decay constants in Table \ref{tab:fP}, and  
the $ \b $ and $ \c $ quark masses in (\ref{eq:mb_MS_mb}) and 
(\ref{eq:mc_MS_mc}), suggest that lattice QCD with 
exact chiral symmetry is a viable framework to study heavy quark physics
from the first principles of QCD.

For systems involving $ \u/\d $ quarks, one may use several quark masses
in the range $ m_{u/d} < m_q < m_{s} $ to perform the chiral extrapolation.
To this end, one may choose a coarser lattice (e.g. $ \beta = 7.0 $),
then it is possible to accommodate a wide range of quark masses
$ m_s/4 < m_q \le m_b $ on the $ 42^3 \times 64 $ lattice,
without significant discretization and finite-size errors.
This study is now in progress. 
Evidently, it has become feasible to treat heavy and light quarks 
as Dirac fermions on the lattice, in lattice QCD with exact chiral symmetry.


\vfill\eject
\flushpar
{\bf Acknowledgement}

\noindent

\bigskip

This work was supported in part by the National Science Council,
Republic of China, under the Grant No. NSC95-2112-M002-005 (T.W.C.),  
and Grant No. NSC95-2112-M001-072 (T.H.H.), and by 
the National Center for High Performance Computation, 
and the Computer Center at National Taiwan University.

\bigskip
\bigskip


\end{document}